\documentstyle[12pt]{article}
\begin{document}
\thispagestyle{empty}

\begin{center}
\LARGE \tt \bf{Non-Riemannian cosmic walls as boundaries of spinning matter with torsion.}
\end{center}

\vspace{1.0cm}

\begin{center} {\large L.C. Garcia de Andrade\footnote{Departamento de
F\'{\i}sica Te\'{o}rica - Instituto de F\'{\i}sica - UERJ

Rua S\~{a}o Fco. Xavier 524, Rio de Janeiro, RJ

Maracan\~{a}, CEP:20550-003 , Brasil.

E-Mail.: GARCIA@SYMBCOMP.UERJ.BR}}
\end{center}

\vspace{1.0cm}

\begin{abstract}
An example is given of a plane topological defect solution of linearized Einstein-Cartan (EC) field equation representing a cosmic wall boundary of spinning matter.
The source of Cartan torsion is composed of two orthogonal lines of static polarized spins bounded by the cosmic plane wall.
The Kopczy\'{n}ski- Obukhov - Tresguerres (KOT) spin fluid stress-energy current coincides with thin planar matter current in the static case.
Our solution is similar to Letelier solution of Einstein equation for multiple cosmic strings.
Due to this fact we suggest that the lines of spinning matter could be analogous to multiple cosmic spinning string solution in EC theory of gravity.
When torsion is turned off a pure Riemannian cosmic wall is obtained.
\end{abstract}

\vspace{1.0cm}

\begin{center}
\end{center}

\newpage

\section{Introduction}
\paragraph*{}

Topological space-time defects usually appears in several distinct geometrical forms such as shells, lines and planes \cite{1,2,3}.  These can be investigated either in the context of Riemannian distributions in General Relativity or in the context of EC - gravity.
In the later case torsion loops in Weitzenb\"{o}ck teleparallel spaces or torsion line defects as spinning cosmic strings have been considered recently by P.S.Letelier \cite{4,5}.
In this short note I shall be considering another type of torsion defect, namely a planar thin wall distribution of orthogonal lines of polarized static spinning particles in linearized EC - gravity.
Linearity is consider here to avoid problems with the square of Dirac $ \delta $-functions.
Following Nitsch \cite{6} spinning matter demands that one consider the more general Riemann-Cartan $ U_{4} $ space-time and therefore Weitzenb\"{o}ck teleparallel $ T_{4} $ solutions are not allowed.

\section{Non-Riemannian planar defects and Spin}
\paragraph*{}

Let us now consider the planar space-time given by

\begin{equation}
ds^{2}=({\omega}^{0})^{2}-({\omega}^{1})^{2}-({\omega}^{2})^{2}-({\omega}^{3})^{2}
\label{1}
\end{equation}
where the basis 1-forms $ {\omega}^{r} $ (r=0,1,2,3) are given by

\begin{eqnarray}
{\omega}^{0} & = & e ^{\frac{F}{2}}dt \nonumber \\
{\omega}^{1} & = & e ^{\frac{H}{2}} dx \nonumber \\
{\omega}^{2} & = & e ^{\frac{H}{2}}dy\\
{\omega}^{3} & = & e ^{\frac{G}{2}} dz\nonumber
\label{2}
\end{eqnarray}
where F,H and G are only functions of z.

Torsion 1-forms are chosen such as

\begin{equation}
\begin{array}{llll}
T^{0} & = & J^{0} {\omega}^{0} \wedge {\omega}^{3} \nonumber \\
\\
T^{1} & = & J^{1} {\omega}^{3} \wedge {\omega}^{1} + J^{2} {\omega}^{0} \wedge {\omega}^{1} \nonumber  \\
\\
T^{2} & = & J^{1} {\omega}^{3} \wedge {\omega}^{2} + J^{2} {\omega}^{0} \wedge {\omega}^{2} \nonumber \\
\\
T^{3} & = & J^{3} {\omega}^{0} \wedge {\omega}^{3} \nonumber \\
\end{array}
\label{3}
\end{equation}

Substitution of (\ref{3}) into Cartan first structure equation
\begin{equation}
T^{a}=d{\omega}^{a}+{{\omega}^{a}}_{b} \wedge {\omega}^{b}
\label{4}
\end{equation}

yields the following connection 1-forms
\begin{equation}
\begin{array}{lllll}
{{\omega}^{0}}_{3} & = & \lbrack J^{0} + e^{\frac{-G}{2}} \frac{F'}{2} \rbrack {\omega}^{0} \nonumber \\
\\
{{\omega}^{1}}_{0} & = & \lbrack J^{2} + e^{\frac{-F}{2}} \frac{\dot{H}}{2} \rbrack {\omega}^{0} \nonumber \\
\\
{{\omega}^{1}}_{3} & = & \lbrack J^{1} + e^{\frac{-G}{2}} \frac{H'}{2} \rbrack {\omega}^{1} \nonumber \\
\\
{{\omega}^{2}}_{3} & = & - \lbrack J^{1} + e^{\frac{-G}{2}} \frac{H'}{2} \rbrack {\omega}^{2} \nonumber \\
\end{array}
\label{5}
\end{equation}
and $ J^{3} = \frac{\dot{G}}{2} e^{-\frac{F}{2}} $.  Where dots mean time derivatives and dashes mean z-coordinate derivatives.  To simplify matters we shall consider that only $ J^{0} $ torsion component and the H(z) component of metric are non vanishing.  This choice of metric is similar to Letelier \cite{7} choice for multiple cosmic strings in Riemannian space since $ d{\omega}^{0} = 0 $ and  c  is a constant.

This hypothesis reduces the connection 1-forms (\ref{5}) to 
\begin{equation}
\begin{array}{lll}
{{\omega}^{0}}_{3} = J^{0}{\omega}^{0}\nonumber \\
\\
{{\omega}^{1}}_{3} = c \frac{H'}{2}{\omega}^{1} \nonumber \\
\\
{{\omega}^{2}}_{3} = -c \frac{H'}{2}{\omega}^{2} \nonumber \\
\end{array}
\label{6}
\end{equation}
since the $ d{\omega}^{0} = 0 $ and c is a constant.

\section{Field equations}
\paragraph*{}

In the language of exterior differential forms the EC - field equations \cite{8,9} are 

\begin{equation}
R^{ik} \wedge {\omega}^{l} {\epsilon}_{ikml} = -16 {\pi}G {\Sigma}_{m}
\label{7}
\end{equation}

\begin{equation}
T^{k}  \wedge {\omega}^{l} {\epsilon}_{ijkl} = -8  {\pi}G S_{ij}
\label{8}
\end{equation}

Where $ R^{ik} \equiv \frac{1}{2} {R^{ik}}_{rs} {\omega}^{r} \wedge {\omega}^{s} $ is the Riemann-Cartan curvature 2-forms, $ {{\Sigma}}_{m} = \frac{1}{6} {{{\Sigma}}_{m}}^{k} {\epsilon}_{krsf} {\omega}^{r} \wedge {\omega}^{s} \wedge {\omega}^{f} $ is the energy-momentum 3-form current, $ {\epsilon}_{ijrs} $ is the Levi-Civita totally skew-symmetric symbol $ S_{ij} $ is the 3-form spin density.  To solve eqns. (\ref{7}) and (\ref{8}) remains to compute the second Cartan structure eqn.

\begin{equation}
{{R}^{i}}_{k} = d{{\omega}^{i}}_{k} + {{\omega}^{i}}_{l} \wedge {{\omega}^{l}}_{k}
\label{9}
\end{equation}
and to compute the KOP matter-spin current

\begin{equation}
{\Sigma}_{i} = {\epsilon} u_{i}u + p({\eta}_{i} + u_{i}u) - 2u^{k} \dot{S}_{ik}u
\label{10}
\end{equation}
(for notation see Ref.[12]) which in the case of a thin cosmic wall can be written as

\begin{equation}
{{\Sigma}}_{i} = {{\Sigma}^{w}}_{i} - 2 u^{k} \dot{S}_{ik}u
\label{11}
\end{equation}
where $ {{\Sigma}_{i}}^{\omega} $ corresponds to the planar thin wall stress-energy tensor $ {{{\Sigma}^{\omega}}_{i}}^{k} $ given by

\begin{equation}
{{\Sigma}^{w}_{i}}^{k} = {\sigma} {\delta}(z) diag(1,1,1,0)
\label{12}
\end{equation}
where $ \delta $(z) is the Dirac $ {\delta} $-function and the plane is orthogonal to the z-direction and $ \sigma $ is the constant surface energy-density.
Since we here deal only with static polarized $ \dot{S}_{ik} $ vanishes and (\ref{11}) reduces to the thin planar wall current.

Substitution of (\ref{6}) into (\ref{9}) yields the components

\begin{equation}
\begin{array}{llll}
{{R}^{0}}_{101}({\Gamma}) & = &  c J^{0} \frac{H'}{2} = {{R}^{0}}_{202}({\Gamma}) \nonumber \\
\\
{{R}^{0}}_{330}({\Gamma}) & = & J^{0'} \nonumber \\
\\
{{R}^{1}}_{212}({\Gamma}) & = & {c}^{2} \frac{H'^{2}}{4} \nonumber \\
\\
{{R}^{2}}_{332}({\Gamma}) & = & - {{R}^{1}}_{331}({\Gamma}) = - \frac{1}{2} (H'' + \frac{1}{2}H'^{2}) \nonumber \\
\end{array}
\label{13}
\end{equation}
where $ \Gamma $ is the Riemann-Cartan connection.

Notice that the component $ {{R}^{0}}_{330} $ has a pure torsional contribution.

Since we are dealing only with linearized EC theory terms such $ H^{'2} $ and $ J^{0}H' $  should be dropped. Substitution of (\ref{11}) and (\ref{13}) into (\ref{7}) yields the following fields equations

\begin{equation}
\begin{array}{ll}
H''(z) = 8 {\pi} G {\sigma} {\delta}(z) \nonumber \\
\\
J^{0'} = \frac{8}{3} {\pi} G {\sigma} {\delta}(z) \nonumber \\
\end{array}
\label{14}
\end{equation}

A simple solution of (\ref{14}) reads
\begin{equation}
H'(z) = 8 {\pi} G {\sigma} {\theta}_{0}(z)
\label{15}
\end{equation}
and
\begin{equation}
J^{0} = \frac{8{\pi}G}{3} {\sigma} {\theta}_{0}(z)
\label{16}
\end{equation}

Here $ {\theta}_{0}(z) $ is the Heaviside step function given by

\begin{equation}
{\theta}_{0}(z)=
\left \{
\begin{array}{ll}  
1  , z < 0 \nonumber \\
\frac{1}{2} , z = 0 \nonumber \\
0  , z > 0 \nonumber \\
\end{array}
\right.
\end{equation}

The second equation in (\ref{14}) tell us that $ {\delta}$-Dirac torsion is not compatible with the thin cosmic wall as far as our model is concerned.

Thus eqn. (\ref{16}) yields a torsion step function. This is not the first time that torsion step functions appear in the context of EC-gravity. Previously H.Rumpf \cite{13} has made use of torsion steps as a mechanism to create Dirac particles on torsion and electromagnetic backgrounds.

\section{Matching conditions}
\paragraph*{}

Equation (\ref{15}) yields the space-time region
\begin{equation}
ds^{2} = dt^{2} - dz^{2} - e^{{\beta}z}(dx^{2} + dy^{2}) \hspace{1.0cm} (z<0)
\label{18}
\end{equation}
where $ {\beta} \equiv 8 {\pi}G{\sigma} $. The resulting space-time is obtained by gluing together \cite{14,15} two space-times across a torsion junction given by a cosmic planar thin wall. One space-time is given by expression (\ref{18}) and the other is given by the Minkowski space-time. Note that the boundary conditions \cite{13}

\begin{equation}
g_{ij} {\vert}_{+} = g_{ij} {\vert}_{-}
\label{19}
\end{equation}

\begin{equation}
n_{k}{{{\Sigma}}_{i}}^{k} - n_{i} \overline{K}_{jkl} \overline{K}^{klj} {\vert}_{-} = 0 
\label{20}
\end{equation}

\begin{equation}
n_{k} {{\Sigma}_{ij}}^{k} {\vert}_{-} = 0
\label{21}
\end{equation}
where $ n_{i} $ is the normal vector to $ z=0 $ plane and the bar over the contortion tensor $ K_{ijk} $ are the projections onto the wall are obeyed.  Eqns (\ref{20}) reduces in the linearized case to $ n_{k} {{{\Sigma}}_{i}^{k}}=0 $. Here the plus and minus signs refer to the RHS and LHS of the cosmic thin wall.  Let us now search for the spin distribution corresponding to Cartan torsion $ T^{0}$.

Substitution of $ T^{0} $ into (\ref{8}) yields the following spin 3-forms

\begin{equation}
S_{13}= - \frac{1}{8{\pi}G} {\theta}_{0}(z) {\omega}^{0} \wedge {\omega}^{2} \wedge {\omega}^{3}
\label{22}
\end{equation}

\begin{equation}
S_{23}= \frac{1}{8{\pi}G} {\theta}_{0}(z) {\omega}^{0} \wedge {\omega}^{1} \wedge {\omega}^{3}
\label{23}
\end{equation}

Notice that the spin distribution (\ref{22}) and (\ref{23}) correspond physically to orthogonal lines of polarized spins along $ z=\mbox{const} < 0 $ hypersurfaces.  Note also that spins do not exist only along the cosmic wall (z=0) but also at the LHS of the wall.

\section{Conclusions}
\paragraph*{}

Note that the resulting space-time is not a pure space-time defect since on the LHS of the cosmic wall the space is not Minkowskian.  Notice also that in the Riemannian limit $ (J^{0} \equiv 0 ) $ the curvature components (\ref{13}) reduce to $ {{R}^{2}}_{332}(\{ \}) = {{R}^{1}}_{331}(\{ \})  =  8 {\pi}G {\sigma} {\delta}(z) $ which represents the Riemannian planar thin wall curvatures.  If the solution here may represent a planar thin domain wall is another story that we will appear elsewhere \cite{14,15}.

Since the choice of the metric (\ref{18}) is the same as the Letelier choice for multiple cosmic strings and since there are lines of spinning particles orthogonal to each other along the cosmic wall it is argued that maybe the lines of spinnings particles could be replaced by spinning cosmic strings.  This idea is also supported by proof of Galtsov and Letelier \cite{16} that the chiral conical space-time arising from the spinning particle solution in (2+1)- dimensional gravity by an appropriate boost is the gravitational counterpart for the infinitely thin straight chiral string.

One could also note that in the case of Letelier \cite{7} solution of plane walls crossed by cosmic strings the only interaction between them is via the metric function H(z) in expressions (\ref{1}) and (\ref{2}).

This fact further support our idea that the lines of polarized spinning particles could be analogous to cosmic strings.  As noted by A.Vilenkin \cite{17} the weak field approximation breaks down at large distance from walls and strings, therefore an exact solution of the problem dealt with here is necessary and will be address in future work.  Finally one may notice that torsion here is constant in one side of the cosmic wall and vanishes on the other. This means that our solution does not describe a torsion wall where torsion is given by $\delta$-Dirac functions.  Another place where constant torsion appears is in the study of torsion kinky in Poincar\'{e} gauge field theory \cite{18}.

\section*{Acknowledgments}
\paragraph*{}

I would like to express my gratitude to Prof. F.W.Hehl, P.S.Letelier and A.Wang for helpful discussions on the subject of this paper. Financial support from UERJ and CNPq. is gratefully acknowledged.

\end{document}